# Directionally Locked Heteroepitaxy with a Structurally Modulated van der Waals Material

Nitish Mathur[1], Guangming Cheng[2], Francesc Ballester[3,4], Gabrielle Carrel[1], Vincent M. Plisson[5], Fang Yuan[1], Jiangchang Zheng[6], Caiyun Chen[6], Scott B. Lee[1], Ratnadwip Singha[1,7], Sudipta Chatterjee[1], Kenji Watanabe[8], Takashi Taniguchi[9], Kenneth S. Burch[5], Berthold Jäck[6], Ion Errea[3,4,10], Maia G. Vergniory[3,11], Nan Yao[2], Sanfeng Wu[12], and Leslie M. Schoop[1*]

1. Department of Chemistry, Princeton University, Princeton, NJ 08544, USA.
2. Princeton Materials Institute, Princeton, NJ 08544, USA.
3. Donostia International Physics Center, 20018 Donostia-San Sebastián, Spain.
4. Department of Applied Physics, University of the Basque Country (UPV/EHU), 20018 Donostia-San Sebastián, Spain.
5. Department of Physics, Boston College, Chestnut Hill, MA, USA.
6. Department of Physics, The Hong Kong University of Science and Technology, Clear Water Bay, Kowloon, Hong Kong SAR.
7. Department of Physics, Indian Institute of Technology Guwahati, Assam 781039, India.
8. Research Center for Electronic and Optical Materials, National Institute for Materials Science, 1-1 Namiki, Tsukuba 305-0044, Japan
9. Research Center for Materials Nanoarchitectonics, National Institute for Materials Science, 1-1 Namiki, Tsukuba 305-0044, Japan
10. Centro de Física de Materiales (CSIC-UPV/EHU), 20018 Donostia-San Sebastian, Spain.
11. Département de Physique et Institut Quantique, Université de Sherbrooke, Sherbrooke, J1K 2R1 Québec, Canada.
12. Department of Physics, Princeton University, Princeton, New Jersey 08544, USA

Corresponding author* (email): lschoop@princeton.edu

**ABSTRACT**: Precise orientation of symmetry-mismatched epilayers on van der Waals (vdW) substrates *via* heteroepitaxy has commonly been achieved through surface treatment processes to accommodate weak interlayer registry and bonding strength, thereby limiting the range of material combinations for heterostructure design. In this study, we investigate the influence of lattice instabilities in a structurally modulated vdW $TaCo_2Te_2$ substrate on the growth and alignment of a symmetry-mismatched bulk $Co_xTe_y$ epilayer using *in situ* heating in a transmission electron microscope (TEM). We show that a Peierls-like lattice instability occurs in $TaCo_2Te_2$ at a transition temperature of ~523 K, which was corroborated by phonon calculations. Post-heat-treated samples reveal a thermally induced surface diffusion process and the dominant lateral growth of the $Co_xTe_y$ epilayer on the $TaCo_2Te_2$ vdW layers, as observed in cross-sectional TEM images. Temperature-dependent selected area electron diffraction (SAED) patterns reveal that the quasi-vdW $Co_xTe_y/TaCo_2Te_2$ heterointerface acquires directional locking by aligning larger interlayer lattice mismatch along the lattice instability axis of $TaCo_2Te_2$, while preserving a strong lattice matching along the orthogonal direction. This heterostructure exhibits precise interlayer registry with one-dimensional lattice incommensuration along the lattice instability axis, resulting from structural distortion to accommodate lattice-mismatch strain. Moreover, the interfacial reconstruction of $TaCo_2Te_2$ back to the distorted phase stabilizes the lattice-locking of the quasi-vdW heterointerface at elevated temperatures. These findings encourage the expansion of material diversity for designing and predicting novel multi-dimensional heterostructures by leveraging lattice instabilities to guide epitaxy.

## INTRODUCTION

Developing methods to interface van der Waals (vdW) materials with three-dimensional (3D) bulk materials *via* heteroepitaxy is of strong interest, which supports multijunction device architectures for future downscaled electronics. [1, 2] Wafter-scale growth of vdW heterostructures has been successfully achieved *via* heteroepitaxy.[3, 4] Here, the lack of strong chemical bonds at the surface of vdW materials successfully accommodates interfacial strain and prevents dislocations caused by the lattice mismatch between interfacing materials.[5] Similarly, the growth of 3D bulk materials on a vdW substrate (or vice versa) occurs *via* quasi-vdW (QvdW) epitaxy. The resulting "hybrid bonding" at the 2D/3D interface consists of weak vdW forces and dangling-bond or electrostatic interactions. [6, 7] In previous studies, well-aligned interfaces have been formed in heterostructures *via* QvdW epitaxy between common 2D materials, such as transition-metal dichalcogenides (TMDs), and 3D bulk metals or semiconductor compounds.[5, 6, 8-12] It makes intuitive sense to interface materials with matching interfacial symmetry to achieve precise epitaxial conditions without rotational misalignment between the epilayer and the substrate.[13-15] Previous studies have shown that the strength of QvdW interfacial interactions in symmetry-mismatched heterostructures dictates whether preferential alignment or rotational disorder relaxes the high-energy heterointerface.[16, 17] Generally, growth dynamics have been controlled by pre-surface treatment of the vdW substrate to promote epilayer-substrate interactions. It involves processes such as creating step edges, plasma treatment, or introducing an amorphous buffer seed layer to achieve lattice-locked heteroepitaxy.[18-21] Strong epilayer crystallinity and interlayer registry during heteroepitaxy are critical for wafer-level material growth.

Beyond weak interlayer vdW forces, strain from interfacing multidimensional (2D/3D) materials can be further accommodated by the inherent flexibility of intralayer atomic bonding.[22] Considering structural flexibility as the key criterion, structurally modulated vdW materials could be suitable growth substrates for accommodating symmetry-mismatched heterointerfaces. Modulated structures are a distinct class of materials characterized by atomic positions that systematically deviate from the ideal periodic lattice in one or more crystallographic directions.[23-25] The deviation from the ideal lattice could be either commensurate, incommensurate, or both with respect to the underlying lattice. These are extensively studied for charge density wave (CDW) phases in vdW materials.[26-29] Above the transition temperature ($T_C$), the low-symmetry modulated structure becomes thermally unstable and deforms into a more stable high-symmetry unmodulated structure. This structural phase transformation is characterized by lattice instability. More commonly, these lattice instabilities in a structurally modulated material near $T_C$ transform into dynamical lattice fluctuations that can persist as short-range order (SRO) even above $T_C$.[30] Previous studies have shown that the short-range lattice distortions in the vicinity of crystalline defects, such as edge dislocations, can influence the growth mechanism and the resulting orientation of the epilayer.[31, 32] Similarly, lattice fluctuations are intrinsic to a modulated material that exhibits both anisotropy and dynamical features of soft phonon modes. The question remains whether this inherent flexibility in vdW-modulated-based heterostructures can enable precise orientation during epilayer growth, or whether thermal and interfacial strain effects hinder orientation lock-in. Additionally, the influence of these lattice instabilities on heteroepitaxy is still poorly understood, as this phenomenon is rarely observed above room temperature in vdW-modulated structures where nucleation and growth occur.[33-36] In this work, we introduce vdW $TaCo_2Te_2$ as a growth substrate that exhibits a modulated structure and lattice instability above room temperature. We analyze the growth of 3D $Co_xTe_y$ epilayer on $TaCo_2Te_2$ nanoflakes above $T_C$ using *in situ* heating in a scanning/transmission electron microscope (S/TEM). The analysis reveals that the large lattice-mismatch axis of the heterointerface aligns with the lattice instability axis of $TaCo_2Te_2$, while the in-plane orthogonal axis shows strong lattice matching. We then compare our results with the isostructural undistorted $TaNi_2Te_2$ system and find that the epilayer growth shows rotational misalignments in the interlayer registry. We show that interactions between anisotropic lattice instabilities and QvdW bonding stabilize a one-dimensional (1D) structural modulation at the $Co_xTe_y/TaCo_2Te_2$ heterointerface and a dominant lateral epilayer growth, thereby facilitating directionally locked QvdW epitaxy.

## RESULTS AND DISCUSSION

**Structurally modulated lattice of $TaCo_2Te_2$.** $TaCo_2Te_2$ is a metallic vdW material with an orthorhombic structure (space group 62, more details of structural refinement in **Table S1-4**) and a sextuple-atomic monolayer[37, 38] with a structural modulation along the *a*-axis (**Figure 1a**) at room temperature (RT). The sample must be heated above RT to obtain a high-symmetry undistorted

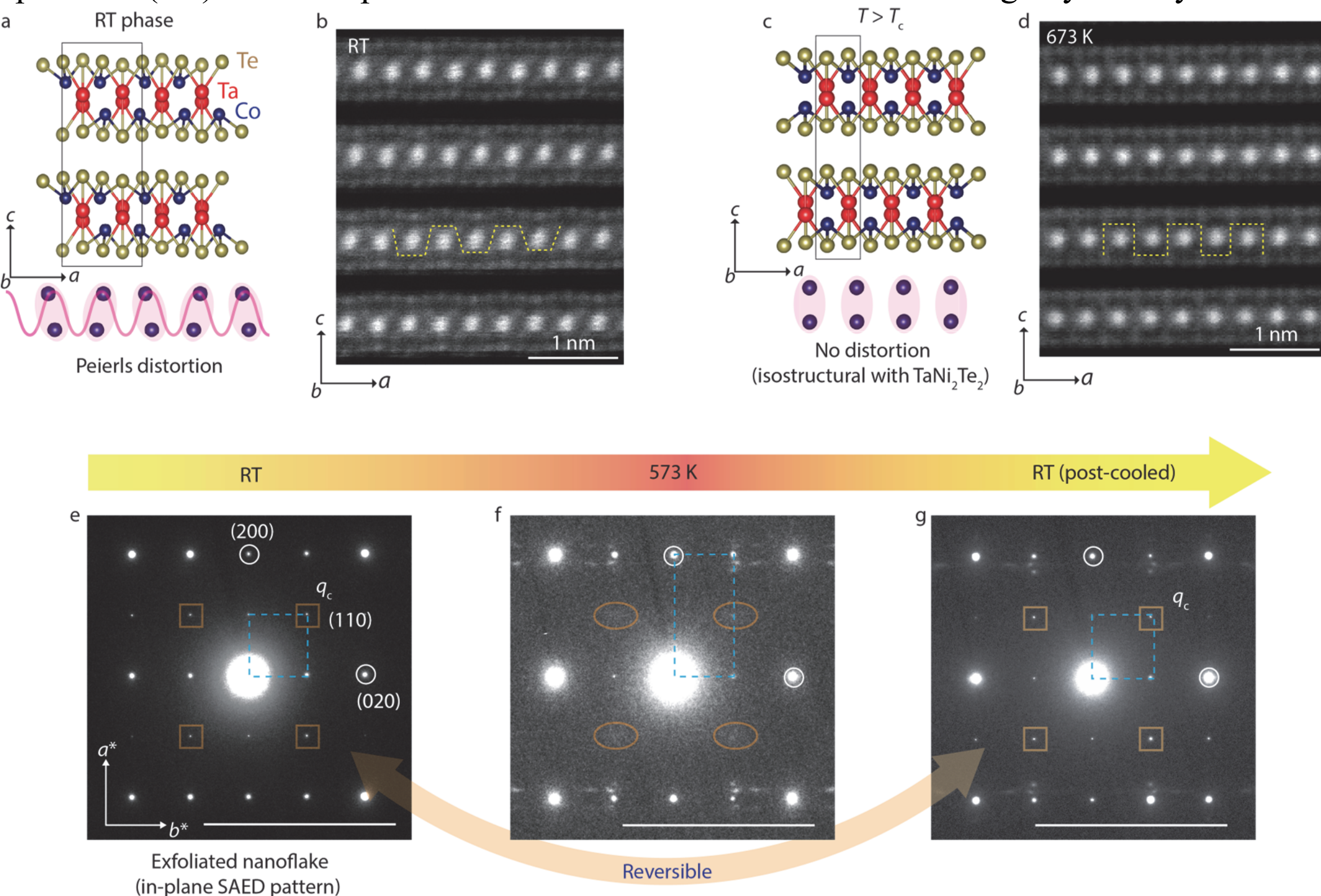

**Figure 1. Structurally modulated $TaCo_2Te_2$. (a-d)** Crystal structure of the room temperature (RT) phase of $TaCo_2Te_2$ with structural modulation, known as Peierls distortion, along the crystallographic *a*-axis and crystal structure of the undistorted phase above $T_C$. Black boxes in panels a and c represent unit cells of both phases. Cross-sectional STEM images of the *ac*-plane showing the transformation from the RT distorted structure to the undistorted structure. Dotted yellow lines highlight Co atomic chains matching corresponding crystal structures in panels a,c **(e-g)** SAED images revealing a reversible structural transition in an exfoliated $TaCo_2Te_2$ nanoflake. Diffraction peaks associated with the Peierls distortion, marked with orange boxes(ellipsoids), indicate the presence(absence) of characteristic commensurate wavevectors ($q_c$) when compared with the lattice parameter of the undistorted reciprocal unit cell marked in dashed blue boxes. Scale bar in panels g-i is 5 $nm^{-1}$. All $TaCo_2Te_2$ SAED images are indexed with respect to the RT distorted structure.

structure with the unit cell halved along the *a*-axis. The $TaCo_2Te_2$ undistorted structure is isostructural with $TaNi_2Te_2$ (**Figure S1**). Using chemical vapor transport (CVT), we synthesize single crystals of $TaCo_2Te_2$ (see Methods**, Figure S2,** and **Table S5**) and find that ultra-thin $TaCo_2Te_2$ nanoflakes from bulk crystals can be easily acquired *via* mechanical exfoliation and do not degrade under ambient conditions (**Figure S3**). We prepare TEM samples by both transferring exfoliated nanoflakes onto a TEM grid and by cutting a thin lamella with a focused ion beam (FIB). Cross-sectional scanning transmission electron microscopy (STEM) images (**Figures 1b,d**) elucidate the transition from the RT distorted structure to the undistorted high temperature structure, as shown in **Figures 1c,d**. We index all $TaCo_2Te_2$ SAED images with respect to the RT-distorted phase. The main reflections of the $TaCo_2Te_2$ RT phase also comprise peaks from the modulated structure (**Figure 1e**), exhibiting commensurate unit cell wavevectors ($q_c$) indexed to the (110) family of planes. We often observe forbidden reflections arising from deviations in the orthorhombic unit cell, which may be caused by residual strains during nanoflake preparation or by strain release *via* stacking faults during sample heating (**Figure S4**). [33, 39]

We determine the temperature-dependent structural evolution of the distorted phase from in-plane SAED images of a $TaCo_2Te_2$ nanoflake. Here, we observe incoherent peaks at $q_c$ above $T_C$ as the structure loses its low-symmetry distorted phase (**Figure 1e**) and transitions to a more stable high-symmetry undistorted phase (**Figure 1f**). Subsequently, the reciprocal unit cell doubles along the $a^*$ direction, which is characteristic of the Peierls-like transition. We fully recover the $TaCo_2Te_2$ distorted phase after cooling from $T_C$ (**Figure 1g**), showing a reversible structural transition. We also detect sharp peaks in differential scanning calorimetry (DSC) scans corresponding to the structural phase transition and obtain a more precise value of $T_C \approx 523$ K (**Figure S5**). Scanning tunneling microscopy (STM) topography acquired at ~ 4.2 K shows the atomically resolved lattice of the $TaCo_2Te_2$ (001) surface (**Figure S6**). Fourier transformed (FFT) image of the topography reveals modulation peaks in addition to Bragg peaks, as in the SAED image at RT. We determine the corresponding real-space modulation ($a_{stripe} \approx 0.67 \pm 0.01$ nm) from the stripe-like pattern by performing an inverse FFT on peaks along the $a^*$ direction. This value is close to the RT lattice parameter ($a \approx 0.66$ nm) of the $TaCo_2Te_2$ distorted structure. Combining the results of TEM diffraction and STM images, we conclude that the modulated structure of $TaCo_2Te_2$ is stable over a wide temperature range.

**Lattice instabilities in $TaCo_2Te_2$.** To understand lattice instabilities during the structural transition, phonon-dispersion calculations for the undistorted structure of $TaCo_2Te_2$ at 0K reveal pronounced imaginary frequencies across multiple high-symmetry points in the Brillouin zone (BZ) (**Figures 2a,b**).[40] The phonon spectra show instabilities at the $X$, $S$, $R$, and $U$ points, which lie along the $k_x$ direction, but also have components along the $k_y$ and $k_z$ directions. However, we found no instabilities at BZ points without a $k_x$ component. This suggests that the unstable modes lie only along the $k_x$ direction and coincide with the Peierls distortion axis in $TaCo_2Te_2$. We found that Co atoms contribute significantly more to this phonon instability than Ta and Te atoms (**Figure S7**). We conduct Raman measurements on an exfoliated $TaCo_2Te_2$ nanoflake (**Figure S8**) within a glovebox[41] at various temperatures to determine changes in collective phonon modes during the structural phase transition. In **Figure 2c**, we show Raman spectra from RT to 600 K, with each temperature offset for clarity. The RT spectrum consists of a cluster of 4 modes from 90 $cm^{-1}$ to 165 $cm^{-1}$ with two additional weak modes between 250 $cm^{-1}$ and 300 $cm^{-1}$. There are two particularly interesting aspects of the temperature-dependent results that we focus on: the sudden change in the spectrum near $T_C$ and the evolution of the P3 and P5 modes over the whole temperature range. While the spectra evolve continuously with increasing temperature, a significant change in the overall spectrum occurs around 550 K, marked by the appearance of a double peak near the P3 mode at 135 $cm^{-1}$. At the same time, the weak P4 mode appears to vanish. These two features become even more significant at the highest temperatures above $T_C$. Beyond the differences in the overall spectra above and below $T_C$, our measurements also reveal an unusual shift in the Raman peaks of the P3 mode. The energy shifts of both these modes are plotted in **Figure 2d**. Typically, phonons shift to lower energies as temperature increases due to lattice expansion. This is precisely how the P5 mode behaves. It gradually softens through P3 mode as it transitions to higher energy levels with increasing temperature. This behavior is indicative of additional higher-order anharmonic terms,[42] most likely due to some avoided band crossings or structural transition above $T_C$. *In situ* heating of an exfoliated $TaCo_2Te_2$ nanoflake above $T_C$ results in a structural transition to the undistorted phase. The persistence of SRO above $T_C$ is evident in the SAED image, which reveals diffuse scattering peaks at $q_c$ (**Figure 3a**). We determine the preferred orientation of incoherent structural domains above $T_C$ by measuring the spread of a representative SRO diffuse peak at the full-width half maximum (FWHM) intensity along the *a** and *b** directions (**Figure S9**). The SRO correlation length is longer along the *a*-axis (≈ 2.12 nm) compared to the *b*-axis (≈ 1.35 nm). Hence, the anisotropic nature of lattice fluctuations is evident in $TaCo_2Te_2$ above $T_C$.

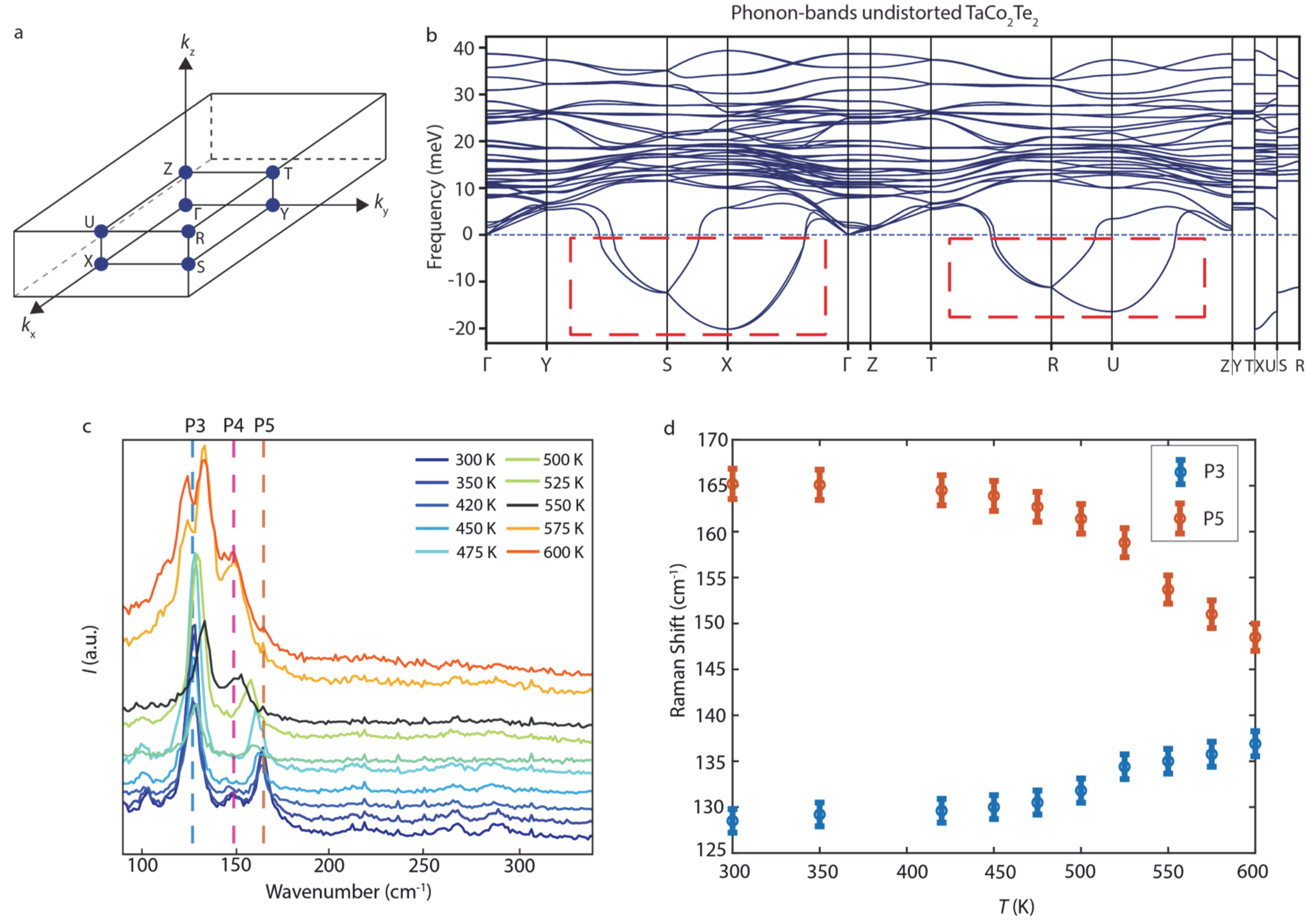

**Figure 2. Lattice instabilities in $TaCo_2Te_2$**. (**a**) 3D Brillouin zone of the undistorted $TaCo_2Te_2$ structure used for phonon-dispersion calculations. (**b**) Phonon band calculations of the undistorted $TaCo_2Te_2$ above $T_C$. Dashed red boxes highlight imaginary phonon bands expanding over a wide region of the Brillouin zone. (**c**) Raman spectrum collected on a $TaCo_2Te_2$ nanoflake at various temperatures above RT. Raman modes P3, P4, and P5 are marked with blue, pink, and orange dashed lines, respectively. (**d**) Corresponding temperature dependence of P3 and P5 Raman peak shifts.

**Emergence of superlattice peaks in $TaCo_2Te_2$ and $TaNi_2Te_2$.** Above $T_C$, we also observe the emergence of the new coherent peaks with increasing temperature in in-plane SAED images of the $TaCo_2Te_2$ exfoliated nanoflake. This manifests as superlattice peaks only along the *a** direction (**Figure 3b**). The persistence of 1D superlattice peaks in the SAED image even after the sample has cooled and rested at RT for an extended time indicates an irreversible phase transition (**Figure 3c**). However, these irreversible superlattice peaks are absent in the SAED image of the post-cooled FIB sample after annealing, where we milled a thickness of 5-10 nm from both the top and bottom (**Figure S10**). Therefore, a new crystalline layer with a different crystal structure forms on the surface of $TaCo_2Te_2$ nanoflakes when heated above $T_C$. We compare the representative superlattice peaks along the *a** direction in SAED images (**Figures 3d,e**) collected at 723 K and at RT (post-cooled). The analysis of SAED images reveals characteristic superlattice peaks appearing at wavevectors $q_1 \approx 0.12\ a^*_{200}$ and $q_2 \approx 0.29\ a^*_{200}$ corresponding (*hk*0) main reflections (where *k* is an odd integer), while their corresponding higher-order wavevectors $\approx 0.24\ a^*_{200}$ ($2q_1$) and ≈ 0.41 $a^*_{200}$ (1– $2q_2$) appear for ($hk$0) main reflections (where $k$ is an even integer or 0) as shown in **Figure 3f**. We also observe weakly coherent peaks in SAED patterns at incommensurate wavevectors that do not orient along the $a^*$ direction (**Figure 3a** and **Figure S11)**. These weakly coherent peaks disappear completely as we increase the temperature above $T_C$, indicating that they

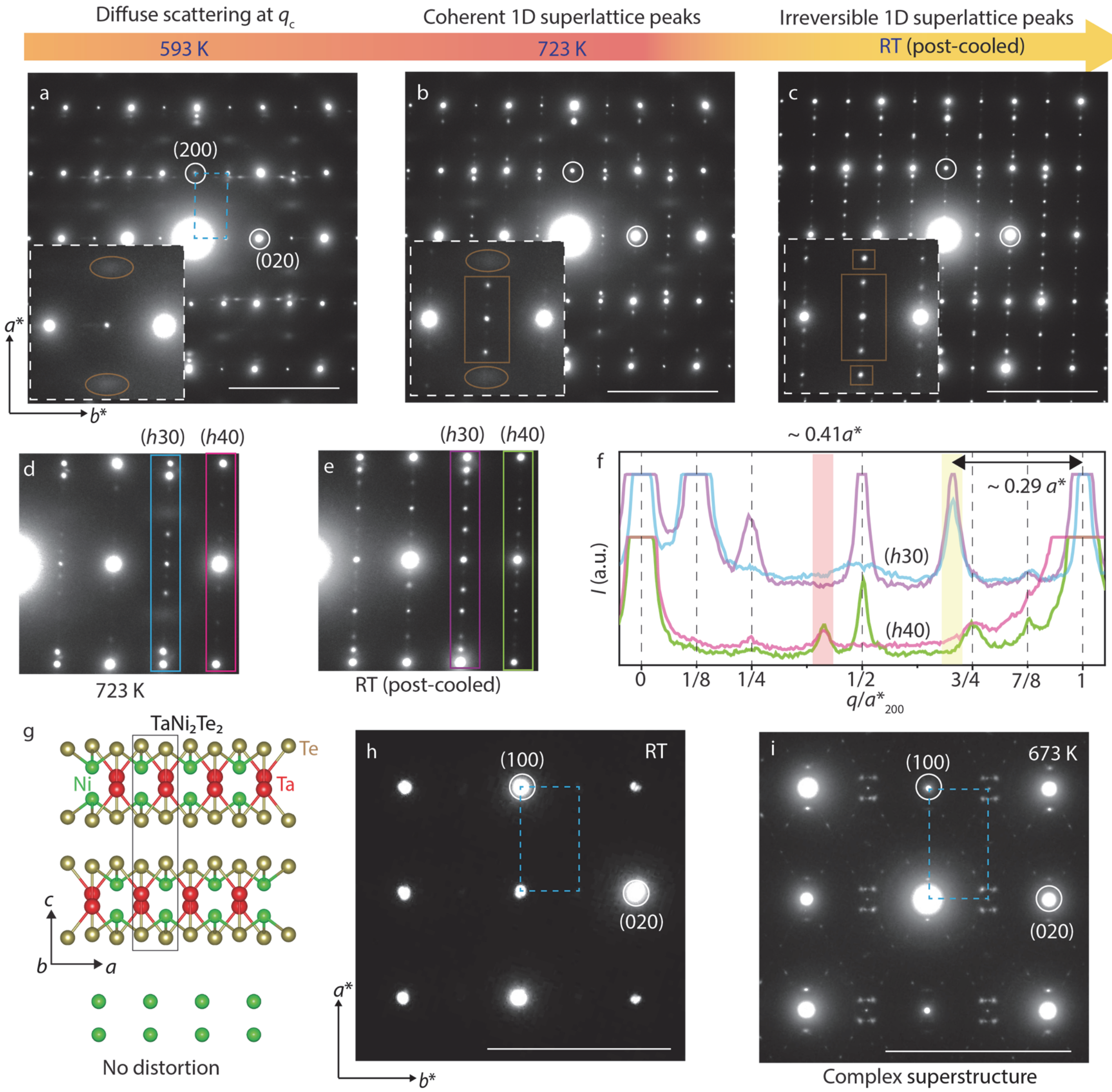

**Figure 3: Emergence of superlattice structure observed using *in situ* heating above $T_C$.** Structure evolution of the $TaCo_2Te_2$ structure above $T_C$ is shown in the in-plane SAED image at (**a**) 593 K, (**b**) 723 K, and (**c**) RT (post-cooled). Insets highlight the corresponding diffuse peaks at $q_c$ associated with SRO, marked by orange ellipsoids, and the coherent 1D superlattice peaks around the main reflections, marked by an orange rectangular box. (**d,e**) Zoomed images of SAED patterns at 723 K and RT (post-cooled). (**f**) Representative intensity line scans along the $a^*$ direction for ($h$30) and ($h$40) planes in the reciprocal space, as shown in panels d and e. The colors of rectangular boxes marked in panels d and e correspond to the colors of trace lines in the intensity plot. **(g)** Schematic of $TaNi_2Te_2$ shows the RT phase, isostructural to the undistorted $TaCo_2Te_2$ phase. SAED patterns show the temperature-dependent structure evolution from the (**h**) RT $TaNi_2Te_2$ phase to a (**i**) complex superstructure at 673 K. Scale bar in panels a-c, h, and i is 5 $nm^{-1}$. SAED images in panels h and i are indexed with the $TaNi_2Te_2$ RT phase.

are linked to kinetic trapping during surface growth. To verify the robustness and alignment of superlattice peaks in $TaCo_2Te_2$, we then conduct *in situ* heating TEM experiments on an in-plane

FIB-prepared lamella, as this sample preparation method differs from that used for exfoliated/transferred $TaCo_2Te_2$ nanoflakes (**Figures S12 and S13**). Indeed, acquired SAED images at increasing temperatures above $T_C$ show similar 1D superlattice peaks with wavevectors ($q_1$, $q_2$) along the $a^*$ direction (**Figures 3d,e**). Note that $q_1$ and $q_2$ are unaffected by decreasing temperature below $T_C$, where the $TaCo_2Te_2$ Peierls distortion occurs. This suggests that the superlattice peaks ($q_1$, $q_2$) are not associated with new $TaCo_2Te_2$ lattice instabilities. To confirm this further, we perform *in situ* heating on exfoliated $TaNi_2Te_2$ nanoflakes (isostructural with the undistorted phases of $TaCo_2Te_2$), which do not show lattice instabilities like $TaCo_2Te_2$ (**Figures 3g,h**).[37] We still observe the emergence of a complex superstructure in the SAED pattern collected at 673 K in $TaNi_2Te_2$ (**Figure 3i**). Despite the isostructural nature of $TaNi_2Te_2$ and undistorted $TaCo_2Te_2$, the superlattice structure differs significantly between them. These observations suggest that the crystalline symmetry of the underlying substrate is not the sole factor determining the interlayer registry of the superstructure.

**Growth of surface layer on $TaCo_2Te_2$.** We examine the thermally induced surface layer (SL) growth using FIB-prepared $TaCo_2Te_2$ cross-sectional TEM samples by sandwiching a $TaCo_2Te_2$ nanoflake between a hexagonal boron nitride (hBN) nanoflake and an amorphous carbon layer (see Methods and **Figure S14**). An atomic-resolution annular dark-field (ADF)-STEM image (**Figure 4a**) near the surface of post-cooled cross-sectional TEM samples (after heating *in situ* at 753 K for 6 hours and then cooling back down to RT) reveals the formation of a 2-3 nm thick new SL. The new SL consists of an ordered SL (OSL) covered by an amorphous SL (ASL). TEM images show that SL uniformly extends laterally across the external surfaces and near the TEM sample's side edges (**Figure S15**). The STEM image contrast decreases in the SL compared to the $TaCo_2Te_2$ vdW layers, which are typically proportional to the atomic number of the constituent elements and the total number of atoms in the atomic column. We used TEM-EDS elemental mapping (**Figures 4b-f**) to investigate this discrepancy, revealing phase segregation in the SL due to surface diffusion at elevated temperatures. This results in Ta atoms migrating to the ASL, whereas Co and Te remain in the OSL. Quantitative analysis using a line scan near the new SL reveals a sharp drop in the Ta relative atomic percentage (%) in the OSL (**Figure 4g**). At the same time, the Co% and Te% remain close to 40%, similar to the $TaCo_2Te_2$ vdW layers. Note that the image drift during the elemental mapping scans does not allow for an exact value of relative elemental composition shown in the line scan plot. Nevertheless, we determine that the Ta-deficient OSL accounts for the decrease in STEM intensity compared to $TaCo_2Te_2$ vdW layers.

Based on these observations, we can elucidate the progression of SL growth in $TaCo_2Te_2$ at temperatures above $T_C$, as shown in the illustration in **Figures 4h-j**. The combined effect of defects, such as vacancies and oxidation-induced amorphization, which are more likely to occur at the sample surface and edges, facilitates thermally induced surface-confined diffusion reactions. TEM-EDS scans reveal that Ta% increases in the ASL layer, where O% is also significantly higher (**Figures 4c,d**). This is expected due to the higher O affinity for Ta.[43] The resulting phase-segregated ASL sublayers are likely to form with high concentrations of Ta-O and Co-Te. Above $T_C$, only Co-Te forms an OSL layer within the operating temperature range and forms a heterostructure with $TaCo_2Te_2$. Similar phase reconstructions have been reported in $GeBi_2Te_4$ following heat treatment, showing that atoms diffuse predominantly in-plane along the vdW gaps and surfaces.[44] In $TaCo_2Te_2$, adatom diffusion within the lateral plane could be facilitated by anisotropic phonon softening above $T_c$, thereby directly guiding the epilayer's nucleation and growth. Furthermore, we observe that the $Co_xTe_y$ epilayer effectively wets the $TaCo_2Te_2$ substrate laterally before vertical growth. The growth style matches the Frank-van der Merwe (FV) model, characterized by layer-by-layer growth and an energetically favorable interface binding energy.[45, 46]

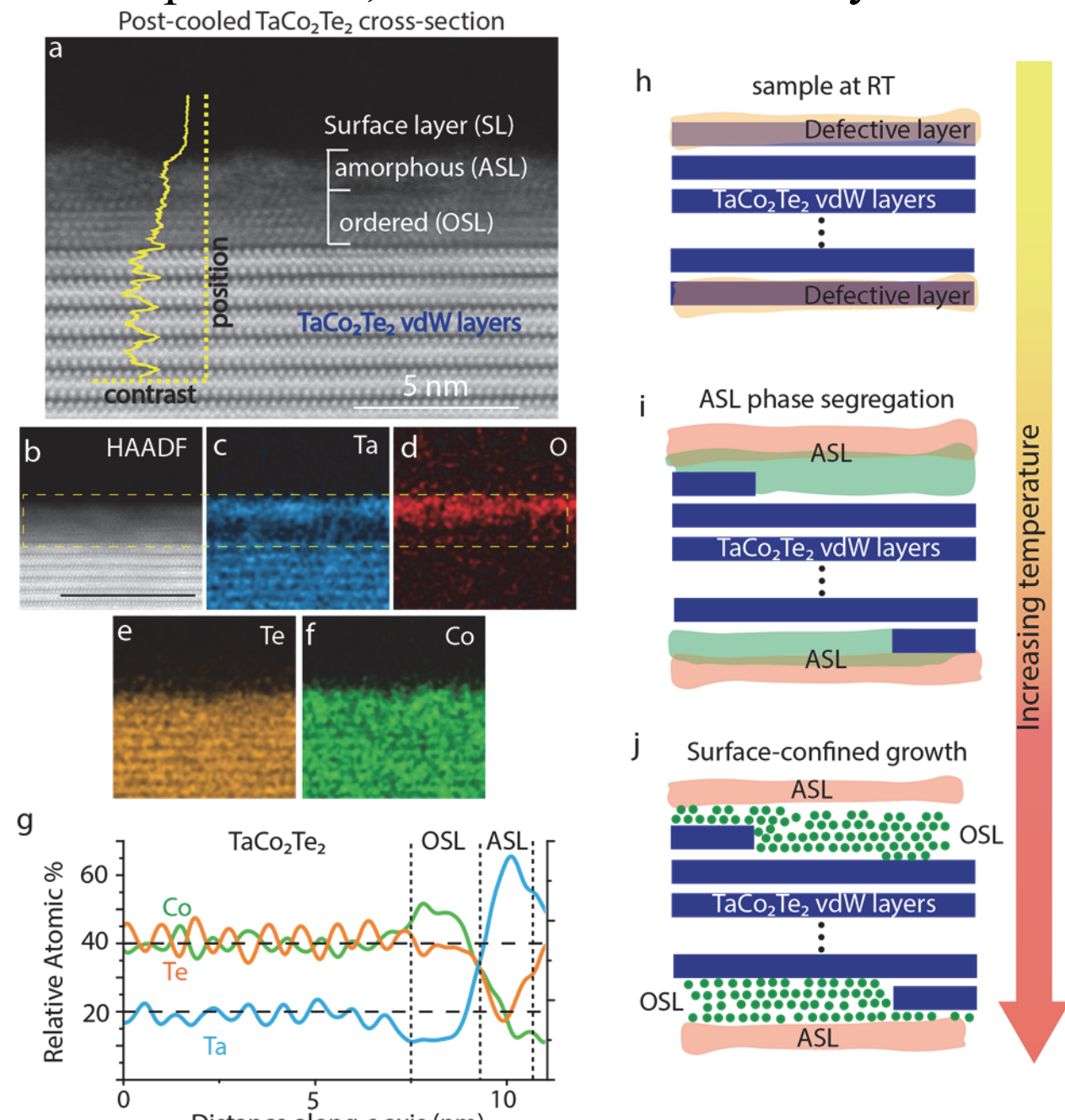

**Figure 4. Growth of surface layer (SL) on $TaCo_2Te_2$ vdW layers.** (**a**) Cross-sectional ADF-STEM image of the post-cooled $TaCo_2Te_2$ FIB-prepared sample showing a SL consisting of amorphous (ASL) and ordered (OSL) layers. The SL layer shows weaker STEM contrast than the $TaCo_2Te_2$ layers. (**b-f**) Elemental mapping of the corresponding cross-sectional HAADF image. The yellow dash box highlighting the Ta and O migration in the ASL layer. (**g**) Quantitative elemental line scan along the *c*-axis marked in a yellow dashed line in panel a. (**h-j**) Schematics illustrating the temperature-dependent structure evolution of SL on $TaCo_2Te_2$ vdW layers.

**Directionally locked heteroepitaxy between the epilayer and the $TaCo_2Te_2$ substrate.** Next, we determine the epitaxial conditions and atomic-level reorganization of the $Co_xTe_y/TaCo_2Te_2$ heterostructure. The $Co_xTe_y$ OSL interface with $TaCo_2Te_2$ at a quasi-vdW gap, as shown in the atomic resolution ADF-STEM image (**Figure 5a**). The crystal structures of $Co_xTe_y$ binary alloys typically adopt a hexagonal or trigonal symmetry, except for an orthorhombic symmetry (*P*nnm) previously reported for $CoTe_2$.[47] In the $Co_xTe_y$ OSL layer, the atomic column aligns along the

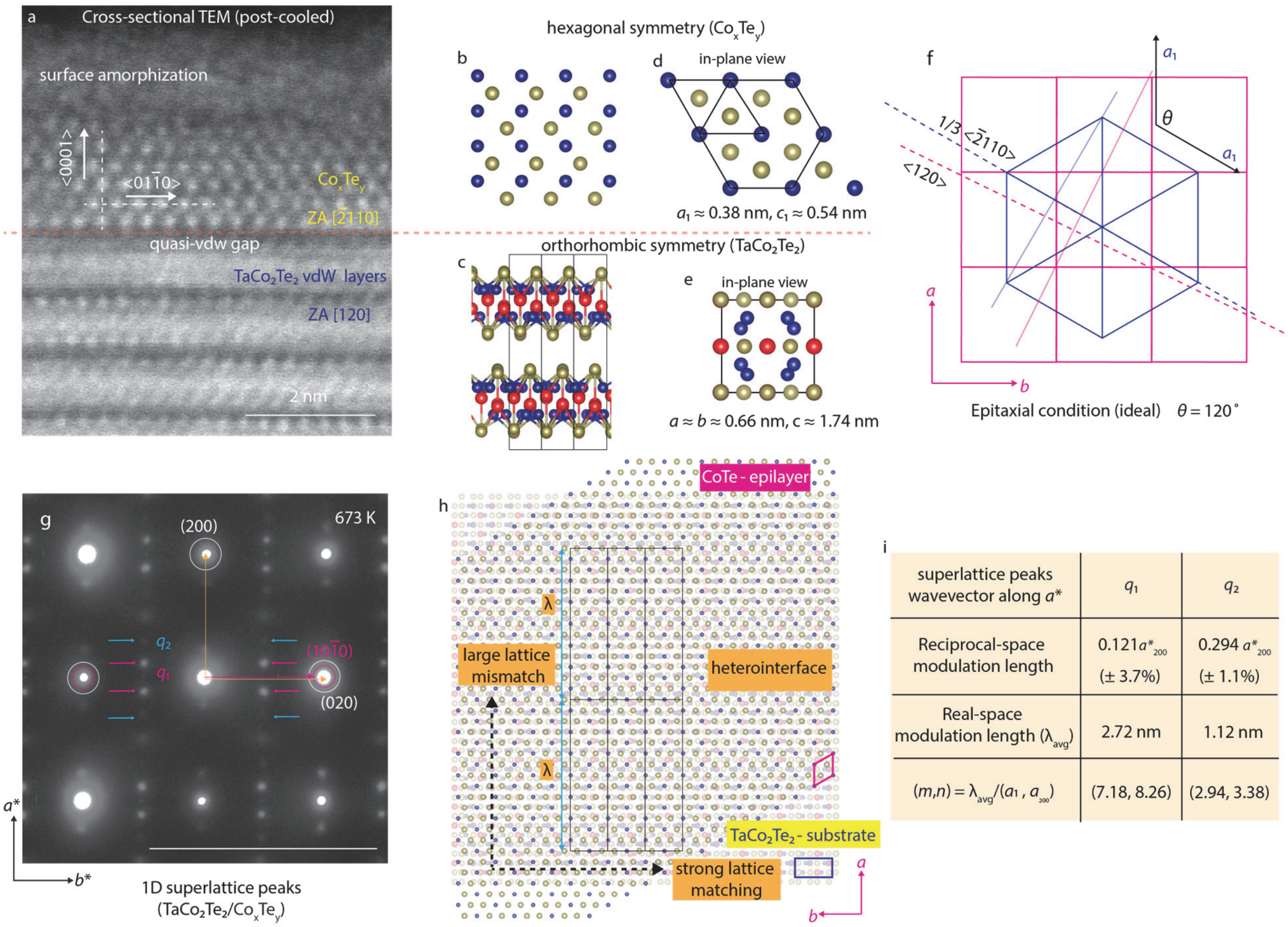

| superlattice peaks wavevector along $a^*$ | $q_1$ | $q_2$ |
| --- | --- | --- |
| Reciprocal-space modulation length | $0.121 a^*_{200}$ (± 3.7%) | $0.294 a^*_{200}$ (± 1.1%) |
| Real-space modulation length ($\lambda_{avg}$) | 2.72 nm | 1.12 nm |
| $(m,n) = \lambda_{avg}/(a_1, a_{200})$ | (7.18, 8.26) | (2.94, 3.38) |

**Figure 5. Structural analysis of QvdW $Co_xTe_y/TaCo_2Te_2$ directionally locked heteroepitaxy. (a)** Cross-sectional atomic resolution ADF-STEM image of the $Co_xTe_y$ /$TaCo_2Te_2$ near the surface of the lamella.(**b**,**c**) Crystal structures with layer stacking viewed along the cross-section of the CoTe/$TaCo_2Te_2$ heterostructure in panel a. In-plane view of the crystal structures, highlighting the mismatched symmetries of the (**d**) hexagonal CoTe and (**e**) orthorhombic $TaCo_2Te_2$. (**f**) Schematic showing the ideal epitaxial condition for the superlattice using a two-dimensional representation of the hexagonal CoTe and orthorhombic $TaCo_2Te_2$ in-plane lattice. Faded blue and pink lines highlight planes imaged in the cross-sectional STEM experiment shown in panel a. (**g**) In-plane SAED pattern of $Co_xTe_y/TaCo_2Te_2$ with 1D superlattice peaks. Pink circles highlight diffraction spots indexed to the hexagonal CoTe (0001) plane overlapping with $TaCo_2Te_2$ diffraction spots marked in white circles. **(h)** Schematic representation of the interlayer registry forming CoTe (hexagonal)/$TaCo_2Te_2$ (undistorted) heterostructure corresponding to the epitaxial condition in panel g. Black, pink, and dark blue boxes highlight superlattice, epilayer-CoTe, and substrate-$TaCo_2Te_2$ unit-cells (**i**) Comparison table of superstructure modulations along the *a*-axis from experimental data and the coincidence site lattice analysis. Scale bar in panel f is 5 $nm^{-1}$.

[−2110] zone axis, where the Te atomic column exhibits brighter STEM intensity compared to the Co atomic column (**Figure 5b)**. In the thinner regions near the edges of the TEM cross-sectional TEM sample, the interconnected chemically bonded Co and Te sublayers are clearly visible in the STEM image (**Figure S16**). This confirms that the 3D hexagonal CoTe non-vdW layered structure (hexagonal, space group no. 194) forms the OSL layer. While the exact $Co_xTe_y$ composition may

vary across the surface layer, $Co_xTe_y$ still adopts a non-vdW hexagonal parent structure with minimal change in lattice parameters.[47] We determine that the atomic columns of $TaCo_2Te_2$ in the cross-sectional STEM image closely align along the [120] zone axis (**Figure 5c and Figure S17**). Hence, the resulting heterostructure aligns along their respective *c*-axes ($Co_xTe_y$ <0001> // $TaCo_2Te_2$ <001>) and should form a heterostructure with mismatched out-of-plane symmetry (**Figures 5d,e**).

The formation of 1D superlattice peaks along the lattice instability axis of the $TaCo_2Te_2$ unit cell shows that the QvdW heteroepitaxy of $Co_xTe_y/TaCo_2Te_2$ favors a specific epitaxial alignment. We show a schematic to illustrate the ideal epitaxial condition of the $CoTe/TaCo_2Te_2$ heterostructure from an in-plane view (**Figure 5f**). This proposed model matches the cross-sectional STEM image shown in **Figure 5a**, where the zone axes of hexagonal $Co_xTe_y$ 1/3<−2110> and $TaCo_2Te_2$ <120> are nearly aligned, with only minor deviations (ideally ≈ 3.4°). The superlattice along the *b** direction is commensurate with the $TaCo_2Te_2$ substrate because $d_{010}$ (CoTe) ≈ $d_{020}$ ($TaCo_2Te_2$), where $d_{hkl}$ is the interplanar distance. This perfect interlayer registry along *b** is likely because the minimal theoretical lattice mismatch (Δ) between these lattice vectors is ≈ 2.2% and forms a lattice-locked axis of the heterostructure. In contrast, a significant lattice mismatch of approximately 17.5% exists along the *a** direction, which coincides with the lattice instability axis of $TaCo_2Te_2$ (**Figure S18**). The average real space modulation length ($\lambda_{avg}$) of the superlattice peaks acquired from the SAED images is ≈ 2.72 nm (for $q_1$) and ≈ 1.12 nm ($q_2$) (**Figure 5g**). We first check the possibility of a 1D Moiré pattern in the $Co_xTe_y/TaCo_2Te_2$ heterostructure to explain the superlattice wavevectors. The deviation between the calculated and experimental modulation lengths is quite large (≈ 24%). Further, we present a schematic of the overlaid atomic structure of orthorhombic $TaCo_2Te_2$-substrate (undistorted) and the hexagonal CoTe-epilayer heterointerface (**Figure 5h**) to determine the deviation from lattice commensurability along the *a*-axis using coincidence site lattice (CSL) analysis.[13] The model states that the CSL sites occur at a periodic distance (λ) along the *a*-axis so that the lattice commensuration can be achieved if $\lambda = ma_1 = na$, where *m* and *n* are integers, and $a_1$ and *a* are the lattice parameters of CoTe and undistorted $TaCo_2Te_2$, respectively (see more details in **Figure S19**). Combining the CSL model analysis and experimental values, we obtain non-integer values of (*m*, *n*) ≈ (7.12, 8.26) for $q_1$ and (2.92, 3.38) for $q_2$, indicating a clear deviation from a commensurate superlattice along the *a*-axis. This weak interlayer registry indicates that it is less energetically favorable to force superlattice commensuration along the *a*-axis. For calculations, we assume that both the CoTe-epilayer and the $TaCo_2Te_2$-substrate lattices are rigid. Although this may not accurately represent our scenario, given that lattice instability exists above $T_c$ and the heterostructure can relax *via* distortion. This explains why we did not observe independent Bragg reflections corresponding to the hexagonal $Co_xTe_y$ (0001) plane (**Figure S20**). While the $Co_xTe_y$ epilayer is directionally locked along the *b*-axis, translational symmetry breaking occurs along the *a*-axis to form a modulated epilayer structure, and $Co_xTe_y$ symmetry transforms from $C_6$ to $C_2$.

The resulting interlayer registry shows an anisotropic interlayer coupling strength with strong coupling forming rigid alignment along the *b*-axis and weak interlayer coupling along the lattice instability *a*-axis. Even with this weaker interlayer coupling, the lattice arrangement maintains precise values of $q_1$ and $q_2$ with minimal deviation across samples (**Figure S21**). Here, the dynamically varying periodic lattice potential along the lattice instability axis could be experienced by adatoms during growth. This anisotropic interlayer coupling results in directionally locked QvdW $Co_xTe_y/TaCo_2Te_2$ heteroepitaxy, which minimizes overall interfacial strain through lattice distortion along the lattice instability axis. Therefore, it prevents interfacial relaxation *via* rotational disorder at elevated temperatures. We elucidate this lattice instability-mediated interfacial stabilization by comparing the growth and interlayer registry of $TaCo_2Te_2$- and $TaNi_2Te_2$-based heterostructures. Because of the isostructural nature of $TaNi_2Te_2$, we expect a similar surface diffusion reaction in which Ta migrates to the surface and forms Ni-Te OSL, as discussed above in **Figure 4**. While we expect the interlayer registry to differ in the $TaNi_2Te_2/Ni_xTe_y$ heterostructure, we observe sample-to-sample variations of the superlattice structure. These variations manifest as diffuse, twinned superlattice peaks and new coherent peaks in SAED images at elevated temperatures ($T > 700$ K), unlike in $TaCo_2Te_2$ (**Figure S22**). Due to the absence of lattice instability and the presence of interlayer symmetry mismatch in the $Ni_xTe_y/TaNi_2Te_2$ heterostructure, the epilayer domains show a rotational degree of freedom to reach an energetically favorable configuration with increasing temperature.

**Interfacial reconstruction stabilizes $Co_xTe_y/TaCo_2Te_2$ heterointerface.** The effects of the lattice instability on the interlayer registry between the $Co_xTe_y$ and $TaCo_2Te_2$ are evident from the increased coherence of $q_c$ peaks from diffuse SRO peaks above $T_C$ in the SAED pattern (**Figures 6a,b,** and other samples in **Figure S23**). This suggests the reentry of the distorted $TaCo_2Te_2$ structure. Analysis of the SAED pattern from the cross-section of bulk $TaCo_2Te_2$ along the zone axis [010], i.e., *ac* plane, shows no evidence of reentrant or superlattice peaks along *a**, which confirms that the reentrant phase is not a $TaCo_2Te_2$ bulk phenomenon (**Figure S24**). Notably, these reentrant peaks maintain coherency even in the second heating cycle of the post-cooled sample as well (**Figure S23**). Based on these observations, we conclude that there exists irreversible surface reconstruction back to RT distorted $TaCo_2Te_2$ above $T_C$, which assists the stability of $Co_xTe_y$ /$TaCo_2Te_2$ heterointerface. The reentrant transition associated with Peierls distortion should result from the changes in the electron-phonon coupling at the $Co_xTe_y$ /$TaCo_2Te_2$ heterointerface. Previous reports on the enhanced Peierls distortion in ultra-thin bismuth films indicate that this

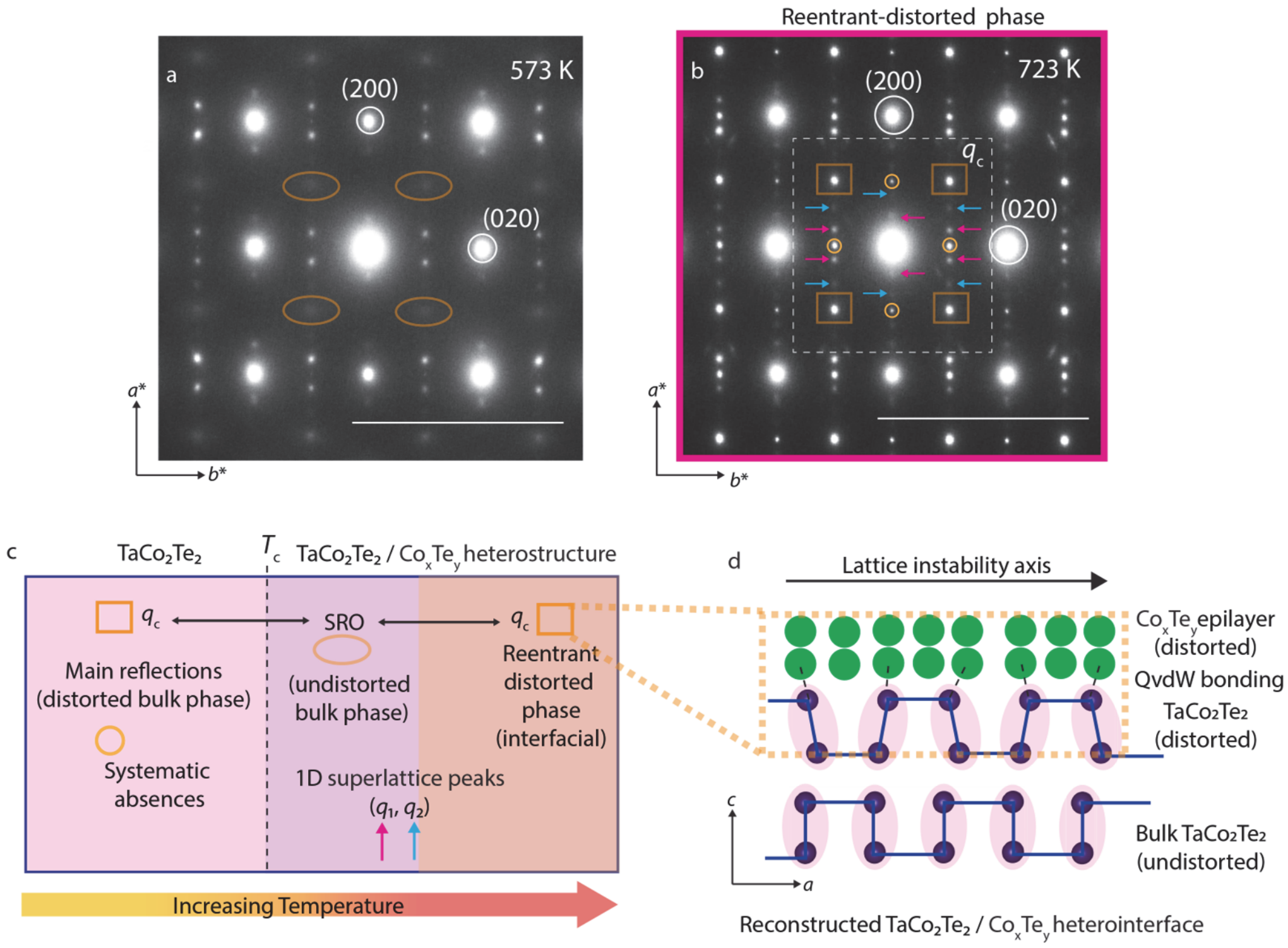

**Figure 6**. **Reconstruction of $TaCo_2Te_2$/$Co_xTe_y$ heterointerface:** (**a-b**) Temperature evolution of $TaCo_2Te_2$ structure above $T_C$ shown in in-plane SAED images. Orange ellipsoids and squares represent SRO and reentrant-distorted peaks, respectively. Dashed white box in panel b, highlighting the systematic absences (orange circles), 1D superlattice (pink and blue arrows), and reentrant-distorted peaks (orange squares). (**c**) Corresponding plot showing the structural phase evolution and emergence of superlattice peaks with increasing temperature. (**d**) Schematic showing the arrangement of the 1D modulated heterointerface with $TaCo_2Te_2$ surface reconstruction, while the rest of the bulk $TaCo_2Te_2$ remains undistorted above $T_C$. Scale bar in panels a and b is 5 $nm^{-1}$.

phenomenon is affected by increased electron localization resulting from bond-confinement effects.[48] We can expect increased localization of electron density of states at the QvdW interface, considering the electronic band structures, where CoTe is a semiconductor, and the undistorted $TaCo_2Te_2$ phase is a metal (**Figure S25**).[38, 47]

We summarize our observations in **Figure 6c**, which presents a schematic of the temperature evolution of diffraction peaks in SAED patterns of $TaCo_2Te_2$ nanoflakes (highlighting bulk and interfacial structural transitions). $TaCo_2Te_2$ stabilizes a Peierls-distorted bulk phase below $T_c$, and we observe corresponding sharp peaks at $q_c$. Above $T_c$, the bulk phase transitions to a more stable undistorted phase and exhibits SRO peaks at $q_c$ associated with anisotropic lattice fluctuations. In addition, we observe weak 1D superlattice peaks ($q_1$, $q_2$) along the lattice instability *a*-axis of $TaCo_2Te_2$, while the superlattice remains commensurate along the perpendicular *b*-axis. The intensities of $q_1$ and $q_2$ increase with temperature. The lattice instability facilitates lateral growth and directional locking of the modulated $Co_xTe_y$ epilayer on the $TaCo_2Te_2$ substrate. Finally, the reentrant peaks at $q_c$ emerge without affecting the coherence of $q_1$ and $q_2$. The interlayer bond interactions at the QvdW $Co_xTe_y/TaCo_2Te_2$ interface along the lattice-locked *b*-axis should alter the $TaCo_2Te_2$ soft-phonon dynamics by suppressing lattice fluctuations. To maintain precise orientation lock-in between $Co_xTe_y/TaCo_2Te_2$, lattice fluctuations induced by increasing thermal energy can be further suppressed by surface distortion. Once sufficient epilayer coverage is achieved at high temperature, the reentrant distorted surface reconstruction peaks at $q_c$ become coherent in SAED images. We finally observe peaks that deviate from the *a*-axis at high temperatures (723-753K) in FIB-prepared samples. However, this does not compromise the coherence and positions of 1D superlattice peaks (**Figure S26**). In principle, the epilayer grown away from the QvdW interface should not experience the same lattice locking effects as the one closer to it. Therefore, rotation misalignment due to thermal relaxation becomes more likely as the epilayer thickness increases. The growth window and thickness for a lattice-locked epilayer using a structurally modulated substrate should be investigated in future studies. Our work encourages studies of predicting novel heterostructures with material combinations in which lattice instabilities can dictate the growth and epilayer orientation.

## CONCLUSIONS

In summary, we show that the lattice instability of a structurally modulated $TaCo_2Te_2$ substrate can be leveraged to achieve robust epitaxial growth and precise alignment of the symmetry-mismatched bulk $Co_xTe_y$ epilayer. The dynamic lattice instabilities predicted by phonon calculations and confirmed through Raman and TEM measurements support anisotropic lattice fluctuations above $T_C$ in $TaCo_2Te_2$. Atomic-resolution ADF-STEM images collected from post-heat-treated cross-sectional samples reveal the growth progression of the crystalline $Co_xTe_y$ surface layer on $TaCo_2Te_2$. To stabilize the high-energy $Co_xTe_y$/$TaCo_2Te_2$ heterointerface, a coherent 1D incommensurate superlattice forms along the lattice-instability axis, while strong lattice locking persists along the other orthorhombic axis. The emergence of the reentrant distorted phase far above $T_C$ further lowers the overall interfacial energy by altering the soft-phonon dynamics of $TaCo_2Te_2$ at the heterointerface. Interfacial reconstruction arising from interactions between substrate lattice instability and QvdW bonding facilitates a structurally modulated interface, resulting in directionally locked heteroepitaxy even at elevated temperatures. Our work presents a strategy for robust multidimensional heteroepitaxy that leverages lattice instabilities to achieve precise interlayer registry by limiting rotational disorders and offers new design rules for deterministic assembly of symmetry-mismatched heterointerfaces.

## METHODS

**Single crystal synthesis and characterization**: Bulk single crystals of $TaCo_2Te_2$ were synthesized using CVT. Synthesis of large single crystals of $TaCo_2Te_2$ was adopted from ref [38]. Oriented single crystals with long axis along the *a*-direction prepared from elemental powder of Ta (Sigma-Aldrich 99.99%), Co (Sigma-Aldrich 99.9%) and Te (Sigma-Aldrich 99.999%), mixed in stoichiometric amount with 30 mg iodine chunks (Sigma-Aldrich 99.9%) and sealed in an evacuated quartz tube, which was heated to 1273 K for 10 days in a box furnace. Acquired mm-sized oriented $TaCo_2Te_2$ single crystals picked from the post-reaction powder mixture using an optical microscope. Similarly, $TaNi_2Te_2$ single crystals were synthesized by replacing Co with Ni (Sigma-Aldrich 99.7%) powder. The sealed quartz tube was placed in a tube furnace for 10 days under a temperature gradient, with the powder mixture maintained at 875 °C at the hotter end and 775 °C at the colder end. After the reaction, $TaNi_2Te_2$ single crystals were collected at the hot and cold ends. The morphology and elemental composition of grown crystals were determined using Quanta 200 FEG ESEM equipped with energy dispersive X-ray (EDX) operating at 15 kV. Single-crystal X-ray diffraction (XRD) measurements were performed on single crystals <100 μm in size using a Bruker D8 VENTURE diffractometer with a PHOTON III CPAD detector and a graphite-monochromated Mo-$K_\alpha$ radiation source. All structure refinements were performed using the OLEX2 software package.

**$TaCo_2Te_2$ nanoflake preparation**: $TaCo_2Te_2$ single crystals were mechanically exfoliated using Scotch Magic tape (3M). Prior to the transfer process, $SiO_2$/Si substrates with a 285 nm oxide layer were thoroughly cleaned by ultra-sonication with acetone, followed by the removal of organic residues using Ar/$O_2$ (45/15 sccm) plasma cleaning (PE-50 PLASMA ETCH INC) for approximately 5 minutes. Following substrate preparation, a high density of laterally micron-sized vdW thin nanoflakes with varying thicknesses was transferred onto the cleaned $SiO_2$/Si substrate. Atomic force microscopy (Bruker Dimension Icon AFM) was performed on thin exfoliated nanoflakes, and images were processed using NanoScope Analysis software. These flakes were picked up in ambient conditions from the $SiO_2$/Si substrate using polymer stamps and a transfer stage equipped with micro-manipulators. Glass-slide stamps were prepared using polycaprolactone (PCL, Sigma Aldrich, $M_n$:80000) solution (10% weight in chloroform), which was spin-coated onto the dome-shaped PDMS. Initially, a TEM Au grid (G2000HAG, Ted Pella INC.) was attached to a bare Si/$SiO_2$ substrate using polycarbonyl (PC) solution. Multiple nanoflakes were then picked up and transferred onto the grid/Si substrate following the process detailed in ref.[49] Subsequently, the entire grid/Si substrate assembly was immersed in a chloroform solution for 20 minutes to dissolve the post-transfer PC and PCL polymers. Finally, the TEM grid was carefully scooped from the chloroform solution and air-dried for 1 min.

**FIB fabricated thin lamella**: TEM thin lamellae were prepared using FIB cutting with a FEI Helios NanoLab™ 600 dual beam system (FIB/SEM). Oriented single crystals of $TaCo_2Te_2$ with their long axis along the *a*-axis were used as a guide to fabricate multiple lamellae for TEM measurements. These lamellae were cut both in-plane (*ab*-plane) and cross-sectionally (*ac*-planes) using FIB techniques. FIB lamella was attached to Mo grid (10GM02, Ted Pella PELCO®) *via in situ* Pt deposition at shared edges. The cross-sectional STEM sample of the $TaCo_2Te_2$ nanoflake was prepared by transferring the nanoflake onto the top of the hBN nanoflake on a Si substrate. Next, a thin amorphous carbon layer was deposited on top of the $TaCo_2Te_2$/hBN assembly. Finally, a cross-section sample was prepared using FIB from amorphous carbon/$TaCo_2Te_2$/hBN/$SiO_2$/Si assembly.

**TEM microscopy:** In-situ heating TEM experiments were conducted using a Gatan double-tilt holder (Model 652) within the TEM column with a base pressure $\approx 10^{-7}$ torr. Temperatures were set and monitored using a Gatan temperature controller (1905). After reaching the set temperature, the electron beam was blanked, and the sample was allowed to equilibrate for at least 20 minutes to minimize thermal drift before starting image collection. *In situ* cooling was achieved by setting the controller to 300 K (27 °C) and waiting overnight (~ 15 hrs) before acquiring images. SAED and atomic resolution HAADF images were collected on a double Cs-corrected FEI Titan Cubed Themis 300 S/TEM. STEM images were processed using the Gatan Microscopy Suite software. The electron diffraction simulations were performed using the CrystalMaker and SingleCrystal software packages. Crystal structure schematics of $TaCo_2Te_2$ prepared using VESTA 3. ImageJ was used to generate intensity line scans of diffraction peaks in SAED images.

**Glovebox Raman measurements**: The air-stability of $TaCo_2Te_2$ exfoliated nanoflakes was determined using an Argon glovebox Raman setup[41]. Nanoflakes were first exfoliated onto a Si substrate in a glovebox. Raman spectra were collected immediately after exfoliation inside a glovebox and again following a 1-hour exposure to ambient conditions. *In situ* Raman heating experiments were conducted on the same flakes in the same system. For both air-stability and temperature-dependent measurements, Raman spectra were acquired using a commercial WiTec Raman microscope. The system consists of a 100X objective lens and a fiber-coupled 532 nm laser set to 300 μW, with spectrometer integration times of 300 seconds. For the heating experiments, a custom heater stage consisting of a small aluminum block, a cartridge heater, and a thermocouple was used to heat the samples to 600 K. To ensure adequate thermal stability, the sample was held at each temperature for approximately 30 minutes before the next spectra were collected.

**STM imaging:** $TaCo_2Te_2$ samples were cleaved using Kapton tape at room temperature inside an ultra-high vacuum (UHV) chamber with a base pressure of $p \approx 2 \times 10^{-10}$ mbar. After cleaving, the samples were directly transferred into the STM head and cooled down to 4.2 K within 2 hours. STM measurements were conducted using a homebuilt STM instrument under cryogenic (4K $\leq$ T $\leq$ 24 K) and UHV (chamber pressure $\approx 2 \times 10^{-10}$ mbar) conditions using a chemically etched tungsten tip which was prepared on a Cu(111) surface through field emission and controlled indentation, as well as calibrated against the Cu(111) Shockley surface state before each set of measurements. Topographies were recorded using constant current ($I$) mode. STM topography of $TaCo_2Te_2$ surface was obtained by using bias voltage ($V_b$) = −5 mV, $I$ = 4 nA.

**First-principles electronic and phonon calculations:** Density functional theory (DFT) calculations were performed to calculate the electronic band structures of $TaCo_2Te_2$ using the Vienna Ab initio Simulation Package (VASP) $v5.4.4$ using the PBE functional. PAW potentials were chosen based on recommended potentials. We used a plane wave energy cutoff of 600 eV. We used $\Gamma -$centered $15 \times 15 \times 5$ and $30 \times 15 \times 5$ $k$-meshes for the undistorted and RT $TaCo_2Te_2$ structures, respectively. The atomic positions of the undistorted $TaCo_2Te_2$ structure were allowed to relax with an energy convergence criterion of $10^{-7}$ eV. A static calculation was performed on each structure using an energy convergence criterion of $10^{-7}$ eV. Gaussian smoothing was applied to the density of states (DOS) calculation. DFT calculations were performed to calculate phonon band structures of $TaCo_2Te_2$ compounds using simulations using the QuantumESPRESSO package and the PBEsol functionals found in the QuantumESPRESSO pseudopotential database[50, 51]. For $TaCo_2Te_2$, an energy cutoff of 65 Rydberg (Ry), together with a $12 \times 6 \times 3$ $k$-mesh for the self-consistent field (SCF) calculations. An energy threshold of $10^{-10}$ Ry was used for the SCF. Density functional perturbation theory calculations were performed using a $2 \times 1 \times 1$ $q$-mesh and an energy threshold of $10^{-20}$ Ry to determine phonon spectra.

**ASSOCIATED CONTENT**

**Supporting Information**

The Supporting Information is available and includes additional figures, tables, and supplementary notes on the detailed structural characterization of the $TaCo_2Te_2$ single crystal, TEM sample fabrication methods, additional atomic-resolution STEM images, *in situ* TEM studies and diffraction analyses on multiple samples, coincidence site lattice analysis, and DFT calculations.

**Notes**

The authors declare that they have no competing interests.

**ACKNOWLEDGEMENTS**

This project is supported by the Gordon and Betty Moore Foundation's EPIQS initiative through award number GBMF9064 (supported L.M.S. and N.M., R.S., S.C., and F.Y.), the Princeton Center for Complex Materials (PCCM) and the NSF-MRSEC program (supports. L.M.S., F.Y. and S.W.) (MRSEC; DMR-2011750) and the Air Force Office of Scientific Research under Grants No. FA9550-24-1-0110 (supports L.M.S., K.S.B., R.S., and V.M.P.). N.M., F.Y., G.C., N.Y., and L.M.S. acknowledge the sample characterization of the Imaging and Analysis Center (IAC) at Princeton

University, partially supported by the Princeton Center for Complex Materials (PCCM) and the NSF-MRSEC program (MRSEC; DMR-2011750). S.W. acknowledges support from Gordon and Betty Moore Foundation's EPiQS Initiative grant no. GBMF11946. B.J. acknowledges support by the Hong Kong RGC (C6033-22G) and the Croucher Foundation (Grant No. CIA22SC02). C.C. acknowledges support from the Tin Ka Ping Foundation. S.B.L. and G.C. are supported by the National Science Foundation Graduate Research Fellowship Program under Grant No. DGE-2039656. Any opinions, findings, and conclusions or recommendations expressed in this material are those of the author(s) and do not necessarily reflect the views of the National Science Foundation. I.E. acknowledges funding from the Spanish Ministry of Science and Innovation (Grant No. PID2022-142861NA-I00) and the Department of Education, Universities and Research of the Basque Government and the University of the Basque Country (Grant No. IT1527-22). M.G.V. received financial support from the Canada Excellence Research Chairs Program for Topological Quantum Matter. M.G.V. and F.B. thank the support from the Spanish Ministerio de Ciencia e Innovacion grant PID2022-142008NB-I00 and the Ministry for Digital Transformation and of Civil Service of the Spanish Government through the QUANTUM ENIA project call - Quantum Spain project, and by the European Union through the Recovery, Transformation and Resilience Plan - NextGenerationEU within the framework of the Digital Spain 2026 Agenda. This project was partially supported by the European Research Council (ERC) under the European Union's Horizon 2020 Research and Innovation Programme (Grant Agreement No. 101020833). K.W. and T.T. acknowledge support from the JSPS KAKENHI (Grant Numbers 21H05233 and 23H02052), the CREST (JPMJCR24A5), JST and World Premier International Research Center Initiative (WPI), MEXT, Japan.

**AUTHORS CONTRIBUTIONS**

N.M. and L.M.S. conceived the project. N.M. and R.S. synthesized the compound. N.M. designed and conducted the exfoliation/nanoflake transfer process. N.M. and G.C. prepared TEM samples and conducted TEM experiments and analysis. G.C. and F.Y. prepared FIB cut samples. J.C. and C.C. performed STM measurements. B.J. provided insights into and supervised the STM measurements. F.B. performed phonon calculations with the help and insight of I.E. and M.G.V. G.C. performed electronic band structure calculations. V.M.P. carried out Raman measurements under the supervision and guidance of K.S.B. S.B.L. performed single-crystal X-ray measurements and data analysis. S.C. carried out DSC measurements and helped reorganize the main text of the manuscript. K.W. and T.T. provided hBN crystals. N.M. analyzed all data with S.W. and L.M.S. supervising. N.M. and L.M.S. wrote the first draft of the manuscript. All authors contributed to a discussion of the results and provided scientific input to the manuscript.